\documentclass[a4paper,fleqn,twocolumn,usenatbib]{mnras}

\usepackage{mathptmx}
\usepackage{ulem}
\usepackage[T1]{fontenc}
\usepackage{ae,aecompl}

\usepackage{graphicx}	
\usepackage{amsmath}	
\usepackage{amssymb}	
\usepackage{indentfirst}

\def\be{\begin{equation}}
\def\ee{\end{equation}}
\def\bea{\begin{eqnarray}}
\def\eea{\end{eqnarray}}

\title[Constraints on Running of Non-Gaussianity]{Constraints on Running of Non-Gaussianity from Large Scale Structure Probes}

\author[Dai \& Xia]{
Ji-Ping Dai,$^{1}$
Jun-Qing Xia,$^{1}$\thanks{E-mail: xiajq@bnu.edu.cn}
\\
$^{1}$Department of Astronomy, Beijing Normal University, Beijing, 100875, China
}

\date{Accepted 2019 November 4. Received 2019 October 28; in original form 2019 September 16}

\pubyear{2019}

\begin{document}
\label{firstpage}
\pagerange{\pageref{firstpage}--\pageref{lastpage}}
\maketitle

\begin{abstract}
In this letter we present constraints on the scale-dependent ``local'' type primordial non-Gaussianity, {which is described by {non-Gaussianity's spectral index}} $n_{\rm NG}$, from the NRAO VLA Sky Survey and the quasar catalog of the Sloan Digital Sky Survey (SDSS) Data Release 6, together with the SDSS Data Release 12 photo-z sample. Here, we use the auto-correlation analyses of these three probes and their cross-correlation analyses with the cosmic microwave background (CMB)  temperature map, and obtain the tight constraint on the spectral index: {$n_{\rm NG}=0.2 ^{+0.7}_{-1.0}$ } ($1\,\sigma$ C.L.), which shows the {first} competitive constraint on the running of non-Gaussianity from current large-scale structure clustering data. Furthermore, we also perform the forecast calculations and improve the limit of $n_{\rm NG}$ using the future Euclid mission, and obtain the standard deviation at 68\% confidence level: {{$\Delta n_{\rm NG}=1.74$ when considering the fiducial value $f_{\rm NL}=3$}}, which provides the complementary constraining power to those from the CMB bispectrum information.
\end{abstract}




\section{Introduction}

The standard inflationary paradigm predicts a flat Universe perturbed by nearly Gaussian and scale invariant primordial perturbations. These predictions have been confirmed by the increasingly precise measurements of the CMB and the large-scale structure (LSS). Since the last decade, the Planck satellite has confirmed that the initial seeds of structure must have been {close to Gaussian} \citep{ade2016planckG}. However, it is difficult to discriminate between the vast array of inflationary scenarios since most of the present constraints on the Lagrangian of the inflaton field have been obtained from measurements of the two-point function, or power spectrum. Therefore, it is natural to study the non-Gaussianity signatures in higher order correlators.

Considering the non-Gaussianity, the bispectrum of gravitational potential $\Phi(\rm k)$ is defined as,
\be
\left\langle\Phi\left(\mathbf{k}_{1}\right) \Phi\left(\mathbf{k}_{2}\right) \Phi\left(\mathbf{k}_{3}\right)\right\rangle \equiv(2 \pi)^{3} \delta_{D}\left(\mathbf{k}_{123}\right) B_{\Phi}\left(k_{1}, k_{2}, k_{3}\right)~,
\ee
where $\delta_{D}\left(\mathbf{k}_{123}\right) \equiv \delta_{D}\left(\mathbf{k}_{1}+\mathbf{k}_{2}+\mathbf{k}_{3}\right)$ is the Dirac delta function, and
\be
B_\Phi(k_1,k_2,k_3) \equiv f_{\rm NL} F(k_1,k_2,k_3)~,
\ee
where $f_{\rm NL}$ is the amplitude of bispectrum, and $F$ encodes the functional dependence on the specific triangle configurations. Here, we mainly focus on the ``local'' shape which comes from the ``squeezed'' triangles {dominantly} ($k_3 \ll k_1,\,k_2$).

{As shown by \citet{grossi2009large} in numerical simulations with non-Gaussian initial conditions of the local kind, the large-scale halo bias can be greatly affected by relatively small values of $f_{\rm NL}$, which provides another way to test the primordial non-Gaussianity using the properties of the LSS clustering data.
{Based on this point, many works used the auto-correlation power spectrums or auto-correlation functions of high-redshift probes} to constrain $f_{\rm NL}$ because the primordial non-Gaussianity can significantly enhance the clustering power at large scales \citep{xia2010primordial,xia2010constraining,xia2011constraints,2013MNRAS.429.2032N,2014MNRAS.441..486K,
2014PhRvL.113v1301L,alvarez2014testing}.
These works obtained comparable results with the limits from the CMB bispectrum.
}

Even though a scale independent $f_{\rm NL}$ has been widely studied in recent works, a scale-dependent $f_{\rm NL}$ is still well-motivated by theoretical predictions of some inflationary models \citep{chen2005running,khoury2009rapidly,byrnes2010scale1,byrnes2010scale,byrnes2011strongly,riotto2011strongly}.
{To denote the running of $f_{\rm NL}$, a non-Gaussianity's spectral index $n_{\rm NG}$ is defined in analogy to the power spectrum spectral index.}
{The first
detailed forecasts on the running of non-Gaussianity were obtained by \citet{sefusatti2009constraining}, and then by other works \citep{becker2012constraining,biagetti2013testing,giannantonio2012constraining}, in which they performed the forecast analysis using the Fisher matrix method. \citet{becker2012first} took the first step and constrained the {non-Gaussianity's spectral index} using the WMAP bispectrum with KSW bispectrum estimator, $n_{\rm NG}=0.3^{+0.8}_{-0.6}$ at 68\% confidence level. \citet{oppizzi2018cmb} extended their work and included additional shapes and running models, and got $n_{\rm NG}=0.4^{+0.8}_{-0.7}$ ($1\,\sigma$ C.L.) for the ``local'' shape considered in this letter. Recently, Planck collaboration published their new constraints on $n_{\rm NG}$ using the same method \citep{akrami2018planck},{ the results give a large error ($\Delta n_{\rm NG}\sim 2$ for a constant prior) on this parameter because $f_{\rm NL}$ constrained by Planck is close to 0}.

The main purpose of this letter is to constrain the running of the ``local'' type non-Gaussianity using the clustering information of LSS probes. Following \citet{byrnes2010scale}, we parameterize the initial bispectrum with the two scalar fields inflationary model, where both fields contribute to the generation of the perturbations.
\bea
\begin{aligned}
&B_{\Phi}\left(k_{1}, k_{2}, k_{3}\right)=2f_{\rm NL}\times\\
&\Bigg[\left(\frac{\sqrt{k_1k_2}}{k_p}\right)^{n_{\rm NG}} P_{\Phi}(k_1)P_{\Phi}(k_2)
+ 2{\rm Perm}\Bigg]~,
\label{bispectrum}
\end{aligned}
\eea
where $k_p$ is the pivot point, and ${P}_{\Phi}$ is the primordial gravitational potential power spectrum. This kind of template arises, for example, from the mixed inflaton-curvaton scenario.

{In contrast to} previous works \citep{sefusatti2009constraining,biagetti2013testing}, we use the real LSS clustering data including radio sources from the NRAO VLA Sky Survey (NVSS) \citep{condon1998nrao} and the quasar catalogue of the Sloan Digital Sky Survey Release 6 (SDSS DR6 QSOs) \citep{richards2008efficient}, as well as the low redshift probe SDSS Data Release 12 photo-z sample (SDSS DR12 PZs) \citep{beck2016photometric}. We constrain on the running of the ``local'' type non-Gaussianity using the auto-correlation power spectra (ACPS) $C_\ell^{qq}$ of these three LSS surveys, and their cross-correlation power spectra (CCPS) $C_\ell^{qT}$ with the CMB temperature fluctuations from the Planck observation.

\section{Formalism}
We {assume that} the collapsed objects form in extreme peaks of the density field $\delta(\mathbf{x})=\delta \rho / \rho$. The statistics of collapsed objects can be described by the statistics of the density perturbation smoothed on some mass scale $M$. Following \cite{loverde2008effects}, the non-Gaussianity probability density function can be obtained by the Edgeworth expansion \citep{bernardeau2002large}, where the non-Gaussianity mass function is
\bea
\begin{aligned}
&\mathcal{N}^{\rm NG}=\frac{d n(M, z)}{d M}=-\sqrt{\frac{2}{\pi}}\frac{\overline{\rho}}{M} e^{-\frac{\delta_c^{2}}{2 \sigma_{M}^{2}}}\left[\frac{d\ln \sigma_{M}}{d M}\times\right. \\
&\left. \left(\frac{\delta_{c}}{\sigma_{M}}+\frac{S_{3} \sigma_{M}}{6}\left(\frac{\delta_{c}^{4}}{\sigma_{M}^{4}}-2 \frac{\delta_{c}^{2}}{\sigma_{M}^{2}}-1\right)\right) +\frac{1}{6} \frac{d S_{3}}{d M} \sigma_{M}\left(\frac{\delta_{c}^{2}}{\sigma_{M}^{2}}-1\right)\right]~,
\end{aligned}
\eea
The redshift dependence is carried by the threshold for collapse $\delta_{c}(z) \approx 1.686 / D(z)$, with $D(z)$ the growth factor. {It worth noticing} that we may replace $\delta_c$ with $\delta_{ec}=\sqrt{q}\delta_c$ for ellipsoidal collapse, and the correction $q=0.75$ results from the N-body simulation \citep{grossi2009large}. $\sigma_M^2$ is the variance of the smoothed density fluctuation, and $S_3$ is the skewness which defines as $S_3={\left\langle\delta_{M}^{3}\right\rangle_{c}}/{\left\langle\delta_{M}^{2}\right\rangle_{c}^{2}}$.
Taking Eq.(\ref{bispectrum}) into the above equation, we can calculate the correction on the Gaussianity mass function when considering the running non-Gaussianity model.

{For the ``local'' type primordial non-Gassianity, there are several theoretical expressions for the large-scale bias \citep{bias4,bias1,bias2,bias3,mcdonald2008primordial,desjacques2011accurate, desjacques2011non, scoccimarro2012large}. In our analysis, we use the accurate prediction for the scale dependent bias correction from primordial non-Gaussianity \citep{desjacques2011accurate}.
\be
\Delta b(M,z,k)=2\frac{\mathcal{F}(M,k)}{W_M(k) M(k)}\left(b_L^{\rm G} \delta_c(z) + \frac{1}{D(z)} \frac{d\ln \mathcal{F}(M,k)}{ d \ln \sigma_M}\right)
\ee
where $W_M$ is a top-hat windows function in Fourier space, and $M(k)={2k^2 T(k)}/({3\Omega_m H_0^2})$, in which $T(k)$ is the transfer function, $\Omega_{m}$ is the current fraction of the matter energy density, and $H_0$ is the current Hubble constant. $b_L^{\rm G}$ is the Gaussian Lagrangian bias and the shape function $\mathcal{F}(M,k)$ can be written as
\bea
\begin{aligned}\mathcal{F}(M,k)=& \frac{1}{4\sigma_{M}^{2}{P_{\Phi}(k)}} \int \frac{d^{3} q}{(2\pi)^3} W_M(q) M(q) W_M(|\mathbf{k}-\mathbf{q}|) \\
 & \times M(|\mathbf{k}-\mathbf{q}|) {B_{\Phi}(k,q,|\mathbf{k}-\mathbf{q}|)}~, \end{aligned}
\label{fnl}
\eea
which includes the dependence of $f_{\rm NL}$ and $n_{\rm NG}$. When setting $n_{\rm NG}=0$, $\mathcal{F}(M,k)$ returns to the constant $f_{\rm NL}$ at large scales as expected. {It worth noticing} that Eq.(\ref{fnl}) is applied to the two scalar fields inflationary model, instead of the single scalar field model discussed in \citet{becker2012first}, because in the single scalar field model there is a very strong degeneracy between $f_{\rm NL}$ and $n_{\rm NG}$. Consequently, we can not obtain reasonable {constraints on the latter} from current LSS observations.}

{Making the standard assumption that halos move coherently with the underlying dark matter, the Lagrangian bias is related to the Eulerian one as $b=1+b_L$. We assume the large scale, linear halo bias for the Gaussian case is \citep{sheth1999large}
\bea
\begin{aligned} 1+b_L^{\rm G}=& 1+\frac{1}{D\left(z_{\mathrm{o}}\right)}\left[\frac{q \delta_{c}\left(z_{\mathrm{f}}\right)}{\sigma_{{M}}^{2}}-\frac{1}{\delta_{{c}}\left(z_{\mathrm{f}}\right)}\right] \\ &+\frac{2 p}{\delta_{{c}}\left(z_{\mathrm{f}}\right) D\left(z_{\mathrm{o}}\right)}\left\{1+\left[\frac{q \delta_{{c}}^{2}\left(z_{\mathrm{f}}\right)}{\sigma_{{M}}^{2}}\right]^{p}\right\}^{-1} \end{aligned}
\eea
where $z_{\rm f}$ is the halo formation redshift, and $z_{\rm o}$ is the halo observation redshift. As we are interested in massive halos, we expect that $z_{\rm f}\simeq z_{\rm o}$. Here, $q=0.75$ and $p=0.3$ account for non-spherical collapse and are a fit result from numerical simulations \citep{scoccimarro2001many}}.



Finally, in order to get the effective bias, we need to integrate the halo mass in the given range relevant to our sources,
\be
b_{\rm eff}^{\rm NG}\left(M_{\min }, z, k\right)=\frac{\int_{M_{\min }}^{\infty} b^{\rm NG}(M, z, k) \mathcal{N}^{\rm NG}(M, z) d M}{\int_{M_{\min }}^{\infty} \mathcal{N}^{\rm NG}(M, z) d M}~,
\ee
{where $b^{\rm NG}(M,z,k)=1+b_L^{G}(M,z)+\Delta b(M,z,k)$ is the scale-dependent bias. This effective bias is dependent on the minimal halo mass which differs for different sources.}

\subsection{Data Analysis}

{Since the Limber approximation is accurate below the level of 10\% at $\ell>10$ where we mainly focus on. It is sufficient for the present analysis.} Here we use the public code CAMB\footnote{http://camb.info} \citep{Lewis:1999bs} and {implement} the Limber approximation to calculate the ACPS and CCPS for LSS surveys \citep{limber1953analysis, PhysRevD.78.043519}:
\be
C_{\ell}^{gg}= \int d z \left[b_{\rm eff}^{\rm NG}(z) \frac{d N}{d z}(z)D(z)\right]^2 \frac{H(z)}{c\eta^2} P\left(k=\frac{\ell+1 / 2}{\eta}\right)~,
\ee
\bea
\begin{aligned}
C_{\ell}^{gT}&=\frac{3 \Omega_{m} H_{0}^{2}T_{\mathrm{CMB}}}{c^{3}\left(\ell+\frac{1}{2}\right)^{2}} \int d z b_{\rm eff}^{\rm NG} (z) \frac{d N}{d z}(z) D(z) H(z)\\
 &~~~~\times\frac{d}{d z}\left(\frac{D(z)}{a(z)}\right)  P\left(k=\frac{\ell+{1}/{2}}{\eta}\right)~,
\end{aligned}
\eea

\begin{figure}
	\centering
	\includegraphics[width=1\linewidth]{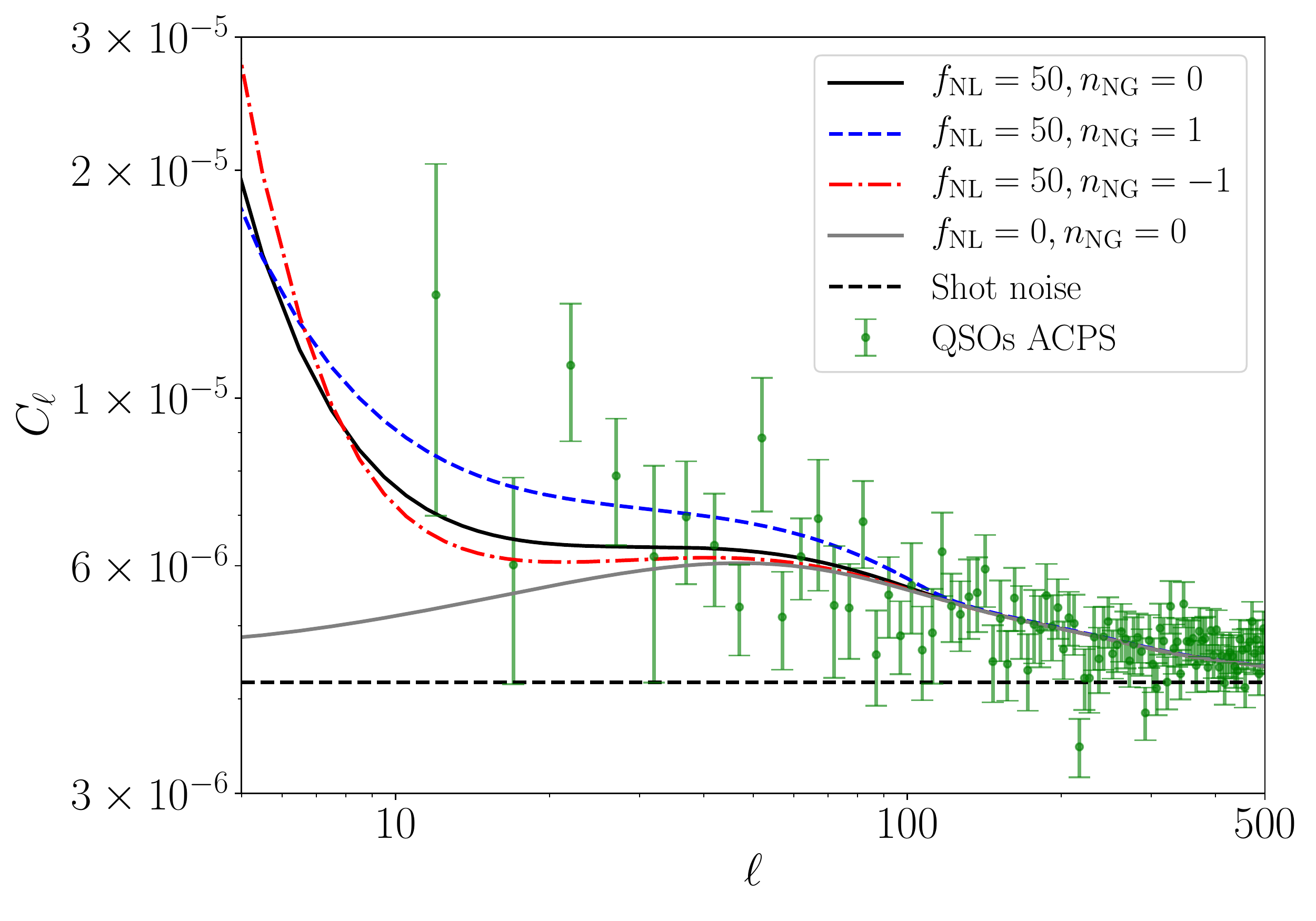}
	\caption{{Observed ACPS of QSO sample, together with theoretical power spectra using different values of the non-Gaussianity and its running. The black dashed line denotes the {shot-noise level in} this QSO catalog.}}
	\label{fig7qsoacf}
\end{figure}

\begin{figure*}
	\centering
	\includegraphics[width=0.4\linewidth]{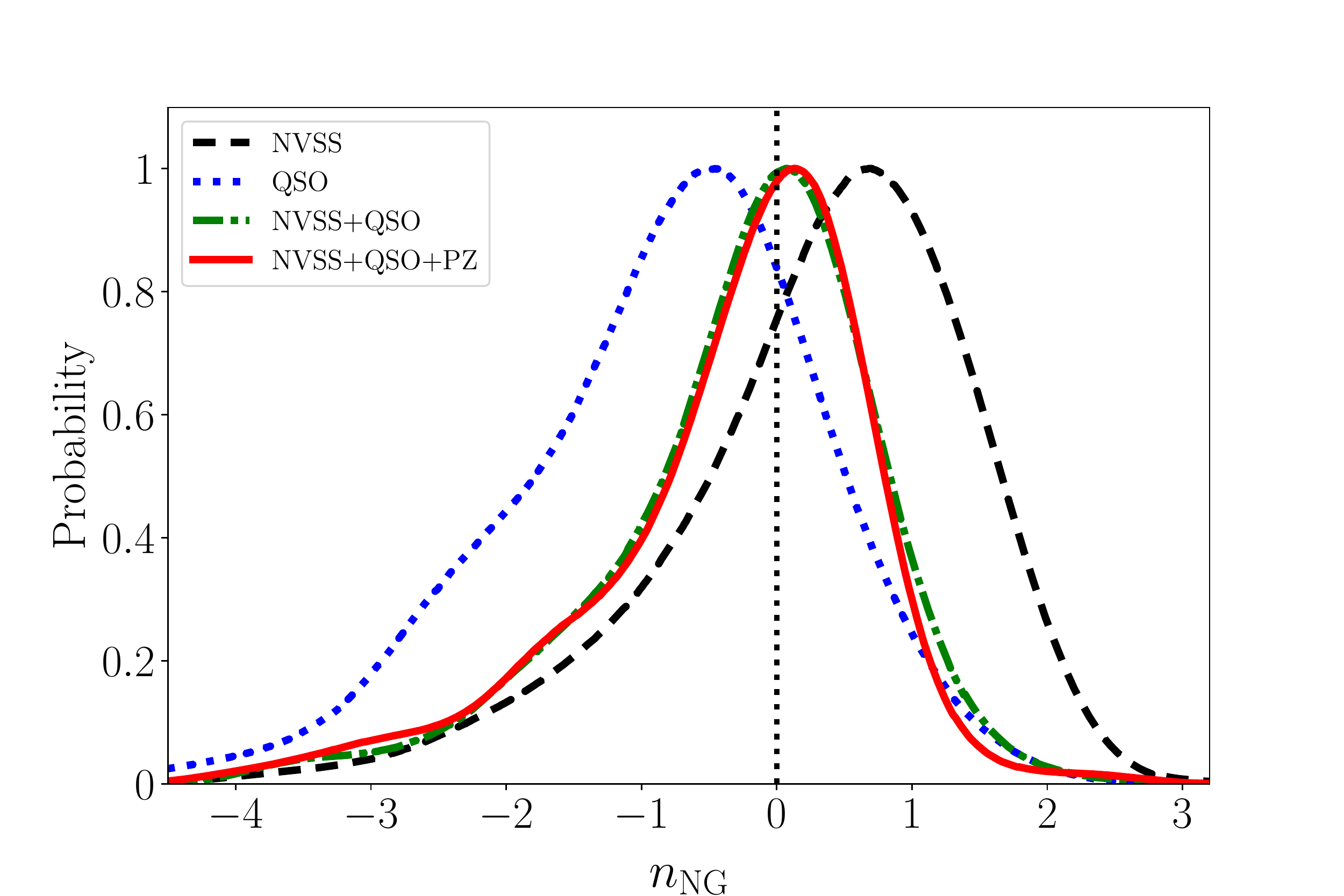}
	\includegraphics[width=0.4\linewidth]{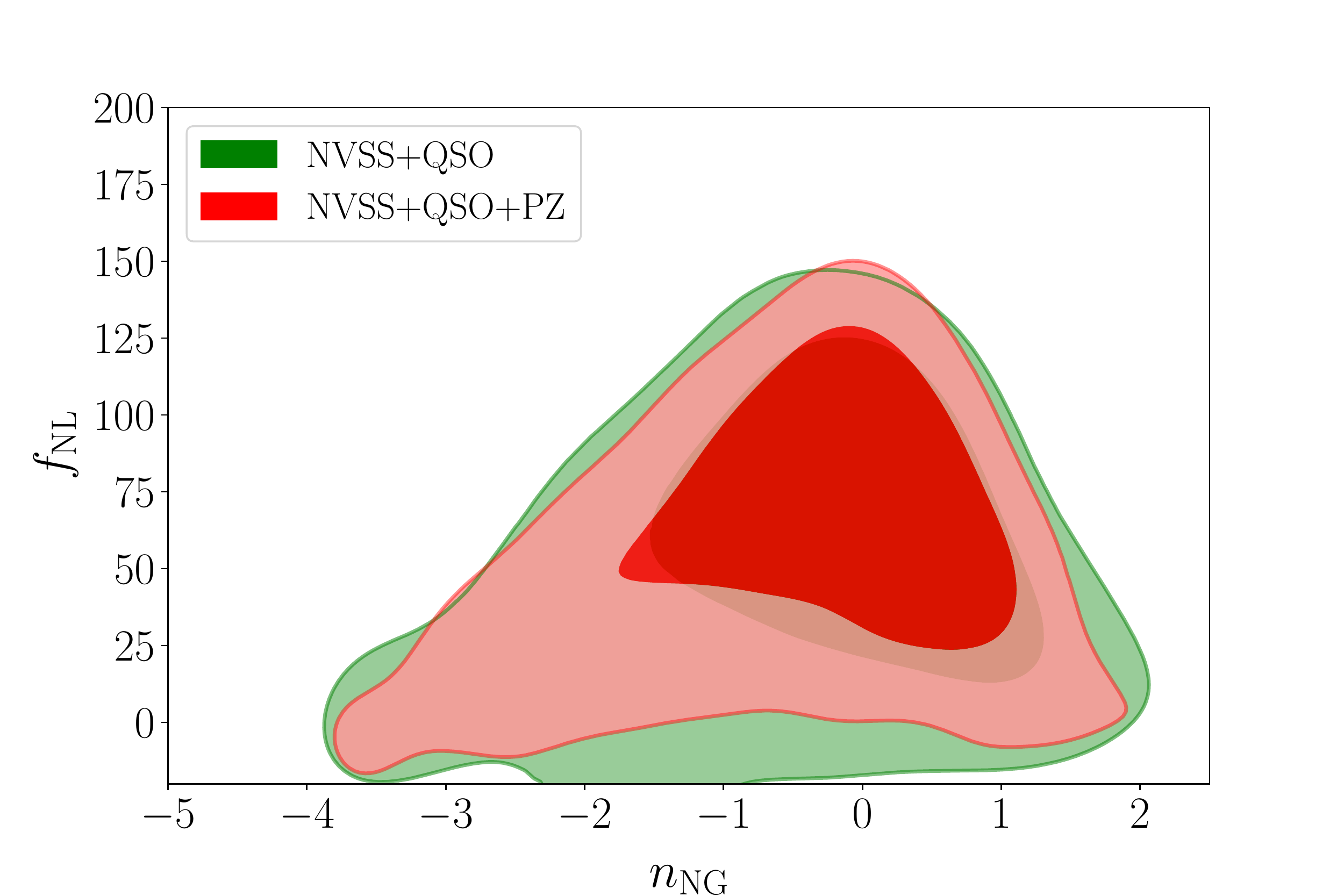}
	\caption{{Marginalized one-dimensional distributions of $n_{\rm NG}$ and two-dimensional distributions ($1,\,2\sigma$ contours) of $n_{\rm NG}$ and $f_{\rm NL}$ using different datasets. See text for more details.}}
	\label{fig10res1}
\end{figure*}

where $dN/dz$ is the normalized selection function of the survey, $\eta$ is the conformal lookback time, and $T_{\rm CMB}$ is the average temperature of CMB photons.

In our work, we use the high redshift probes, NVSS and SDSS DR6 QSOs, together with the low redshift probe SDSS DR12 PZs. {Their redshift ranges span $0<z<3.5, 0<z<5, 0<z<1$, respectively, and we divide them into 200 bins uniformly to calculate the ACPS and CCPS.}  We refer to the recent work \citep{Cuoco:2017bpv} for more details of these samples, {including redshift distributions and masks.} {It is worth mentioning
that several} new quasar catalogs have been published based on the SDSS data release, complemented in some cases with additional information. \citet{Cuoco:2017bpv} has checked the adequacy of these different QSO samples, and showed that all these new samples were detected large variations in the number density of sources across the sky. Therefore, here we still rely on the SDSS DR6 QSOs catalog. The catalog of extragalactic objects are 2D pixelized maps of $n(\hat{\Omega}_i)$, with $N_{\rm side}=512$. We can use PolSpice\footnote{http://www2.iap.fr/users/hivon/software/PolSpice/} \citep{szapudi2001fast, chon2004fast, efstathiou2004maximum, challinor2005error} to estimate the power spectra. Considering the ACPS, the shot-noise should be taken {into account; it is constant} in multipole and can be expressed as $C_N=4\pi f_{\rm sky}/N_{\rm gal}$, where $f_{\rm sky}$ is the fraction of sky covered by the catalog in the unmasked area and $N_{\rm gal}$ is the number of catalog objects in the unmasked area.

In order to show the effects of non-Gaussianity and its running, in Fig. \ref{fig7qsoacf} we present both the theoretical and observed (green points) ACPS, together with shot noise of DR6 QSOs sample (black dashed line). As we can easily see, when comparing the Gaussian case (grey solid line), a positive non-Gaussianity (black solid line) can significantly enhance the clustering power at large scales {($\ell < 60$)}. When including the non-zero {non-Gaussianity's spectral index} $n_{\rm NG}$, the behaviors of ACPS at large scales become different.{  A positive $n_{\rm NG}$ obviously {raises} the amplitude of ACPS at scales $10<\ell<60$ as shown in the blue dashed line, while it plays the opposite way when  $n_{\rm NG}$ is negative (red dash-dotted line).} This discrepancy is the reason why we could use the LSS clustering data to constrain the non-Gaussianity and its {non-Gaussianity's spectral index}.

\begin{table*}
	\caption{{Constraints on $n_{\rm NG}$ and $f_{\rm NL}$ from different datasets, together with constraints from WMAP bispectrum.}}
	\begin{tabular}{c |ccc | ccc }
		\hline
		\hline
		& \multicolumn{3}{c}{$f_{\rm NL}$} & \multicolumn{3}{|c}{$n_{\rm NG}$} \\
		\hline
		&best fit & 1$\sigma$ & 2$\sigma$ &best fit & 1$\sigma$ & 2$\sigma$ \\
		\hline
		NVSS ACPS &51& [10, 106] &  [-53, 144] &0.8& [-0.9, 1.8] &[-2.9, 2.5]\\
		NVSS ACPS+CCPS &53& [14, 103] &  [-31, 144] &0.9& [-0.7, 1.8] &[-2.8, 2.4]\\
		\hline
		QSO ACPS&43 &[-7, 108] & [-84, 135] &-0.7& [-2.1, 0.7] &[-3.8, 1.4] \\
		QSO ACPS+CCPS &41 &[-4, 107] & [-82, 134] &-0.6& [-1.9, 0.6] &[-3.6, 1.4] \\
		\hline
		NVSS+QSO ACPS+CCPS& 54 & [28, 101] & [-18, 132] &0.1 & [-0.9, 0.9] &[-2.5, 1.6]\\
		\hline
		NVSS+QSO+PZ ACPS+CCPS& 58& [31, 103] & [-13, 133] &0.2 & [-0.8, 0.9] &[-2.4, 1.4]\\
		\hline
		WMAP7 \citep{becker2012first}& --- & --- & ---& 0.3 &[-0.3, 1.1] &[-0.9,2.2]\\
		\hline
		WMAP9 \citep{oppizzi2018cmb}& --- & --- & ---& 0.4 & [-0.3,1.2] &---\\
		\hline
	\end{tabular}
	\centering
	\label{tab:1}
\end{table*}

\section{Fitting Results}
In our calculations, we use the public software CosmoMC \footnote{http://cosmologist.info/cosmomc/} \citep{lewis2002cosmological}, a Markov Chain Monte Carlo (MCMC) code to perform the global constraints on $f_{\rm NL}$ and $n_{\rm NG}$ from the ACPS and CCPS clustering data at scales $10<\ell<500$ of three LSS surveys described above. {Here, we abandon the very large scale data mainly for three reasons: 1) to avoid the deviation from accurate theory when we use Limber approximation. 2) to reduce the systematic error at very large scales. 3) to avoid the effects of gauge corrections on the power spectrum on very large scales \citep{yoo2009new}.
A simple $\chi^2$ is used for the fit in our analysis:
\be
\chi^2=\sum_{i\rm th~data} \sum_{\ell} (\hat{C^i_\ell}-C^i_\ell)^T\Gamma_i^{-1}(\hat{C^i_\ell}-C^i_\ell) ,
\ee
where $\Gamma_i$ are the covariance matrixs output from PolSpice estimator and $i$ means different ACPS or CCPS when we preform the the joint analysis. $\hat{C^i_\ell}$ and $C^i_\ell$ represent the model and the measured power spectrum.
Furthermore, we also constrain three minimal halo masses $M_{\min}$ for three LSS surveys, and the constraint results on $M_{\min}$ using their ACPS and CCPS are $10^{12.5\pm0.09}h^{-1}M_{\odot}$ for NVSS, $10^{12.1\pm0.1}h^{-1}M_{\odot}$ for SDSS DR6 QSOs and $10^{11.5\pm0.21}h^{-1}M_{\odot}$ for SDSS DR12 PZs, respectively (1$\sigma$ C.L.).
In order to accelerate the calculation, we do not include the basic cosmological parameters. We admit there are degeneracies between some of the cosmological parameters with $f_{\rm NL}$, but they are constrained very tight by Planck. We have checked our work, and the results show these parameters have very little influence on the final constraints.}
Therefore, we assume the standard $\Lambda$CDM model, with purely adiabatic initial conditions and a flat Universe, and fix the six cosmological parameters as best fit values from the Planck measurement \citep{aghanim2018planck}: $\Omega_b h^2 = 0.0224,~\Omega_c h^2 = 0.1201,~100\theta_{MC} = 1.0409,~\tau = 0.0543,~n_s = 0.9661,~{\rm{ln}}(10^{10}A_s) = 3.0448$.

\citet{oppizzi2018cmb} showed the dependence of the likelihood on the pivot scale $k_p$. As \citet{becker2012first} proposed, the true pivot scale favored by the data is the value of $k_p$ for which the errors in $f_{\rm NL}$ and $n_{\rm NG}$ are uncorrelated. 
In our analysis, we start with an arbitrary value of $k_p$, compute the likelihood and then rescale $k_p$ by \citep{shandera2011generalized},
\be
k_{p}=k_{p}^* \exp \left(-\frac{C_{f_{\rm NL}^*, n_{\rm NG}^*}}{f_{\rm NL}^*C_{{n_{
\rm NG}^*},n_{\rm NG}^*}}\right)~,
\ee
where $k_p^*$ is the arbitrary pivot used initially and $f_{\rm NL}^*$ is the constraint result using $k_p^*$, $C$ is the covariance matrix between $f_{\rm NL}^*$ and $n_{\rm NG}^*$. In practice, we find $k_p=0.016 \rm Mpc^{-1}$ is appropriate in our analysis, since the degeneracy between $f_{\rm NL}$ and $n_{\rm NG}$ we obtain is small enough if using this pivot scale.

We start with the NVSS catalog. In Tab. \ref{tab:1} we list constraints on $n_{\rm NG}$ and $f_{\rm NL}$ from NVSS. If only using the NVSS ACPS, we obtain the constraint: {$f_{\rm NL}=51^{+55}_{-41}$ ($1\,\sigma$ C.L.)}, which is consistent with the Gaussian case at $2\,\sigma$ confidence level, {similar with previous works}. The {non-Gaussianity's spectral index} can also be constrained by the ACPS data: {$n_{\rm NG} = 0.8_{-1.7}^{+1.0}$ }at 68\% C.L., which is consistent with zero at $1\,\sigma$ confidence level, and imples that a positive value is slightly preferred, since in the analysis we find that there is a mild enhancement on the amplitudes of ACPS data points at scales $\ell < 30$. When we combine the ACPS and CCPS of NVSS together, the constraint on $n_{\rm NG}$ is slightly tightened: {$n_{\rm NG} = 0.9_{-1.6}^{+0.9}$} at $1\,\sigma$ confidence level, as shown in the black dashed line in Fig. \ref{fig10res1}. Apparently, in this analysis the constraining power of ACPS is much stronger than the CCPS.

Then we move to the SDSS DR6 QSOs data. Similar with the NVSS result, using ACPS data alone the constraint of non-Gaussianity is consistent with zero safely: {$f_{\rm NL}=43^{+65}_{-50}$} ($1\,\sigma$ C.L.). This result is different from some previous works \citep{xia2011constraints}, which might due to the more conservative mask we use in the analysis. We also obtain the constraint on the {non-Gaussianity's spectral index}: {$n_{\rm NG}=-0.7_{-1.4}^{+1.4}$ }($1\,\sigma$ C.L.), which is also consistent with zero at $1\,\sigma$ confidence level. {Unlike} the NVSS constraint, the DR6 QSO sample slightly {prefers} a negative value of $n_{\rm NG}$, since the ACPS data points at scales $10 < \ell< 100$ are slightly suppressed comparing with the non-running case
. Again, we combine the ACPS and CCPS of SDSS DR6 QSOs, the blue dotted line shows that the constraint only has a very minor change: {$n_{\rm NG} = -0.6_{-1.3}^{+1.2}$} at 68\% confidence level. The non-running case is still favored by the QSO data.

If we combine the ACPS and CCPS data of NVSS and SDSS DR6 QSOs samples together, we obtain tight constraint on the {non-Gaussianity's spectral index}: {$n_{\rm NG} = 0.1_{-1.0}^{+0.8}$ } ($1,\,\sigma$ C.L.), as shown in the green dash-dotted line, which shows a {constraint competitive to previous work \citep{becker2012first, oppizzi2018cmb} that used }the CMB bispectrum to constrain the running non-Gaussianity. In the right panel of Fig. \ref{fig10res1}, we also show the two-dimensional contours between $f_{\rm NL}$ and $n_{\rm NG}$. Clearly we can see that now the degeneracy between them is not very strong, since we set a proper scale $k_p$ in the calculations. Furthermore, we also see a long tail at lower values of $n_{\rm NG}$, because in our calculations we only use the data points at $\ell > 10$. The effect of the negative $n_{\rm NG}$ is smaller than that of the positive {non-Gaussianity's spectral index} at scales $10 < \ell < 100$, as shown in Fig. \ref{fig7qsoacf}. Therefore, the constraining power on the negative $n_{\rm NG}$ will be weaker than that on the positive index. Finally, we add the ACPS and CCPS of SDSS DR12 PZs into the analysis, and {obtain a tight} constraint on the {non-Gaussianity's spectral index} of the ``local'' type non-Gaussianity:
\be
{n_{\rm NG} = 0.2 ^{+0.7}_{-1.0}~~(1\,\sigma~{\rm C.L.})~,}
\ee
which clearly shows that the non-running case is supported by the current LSS clustering data.

{Here we also preform a simple forecast }using the future Euclid observation to estimate constraints on $f_{\rm NL}$ and $n_{\rm NG}$ based on the Fisher matrix technique. 
The Fisher matrix for the running non-Gaussianity parameters $f_{\rm NL}$ and $n_{\rm NG}$ is given by \citep{biagetti2013testing}
\be
\mathcal{F}_{i j}=V_{{\rm surv} } f_{{\rm sky}} \int \frac{\mathrm{d} k k^{2}}{2 \pi^{2}} \frac{1}{2 P_{g}^{2}} \frac{\partial P_{g}}{\partial \theta_{i}} \frac{\partial P_{g}}{\partial \theta_{j}}~,
\ee
where $\theta_i$ are $f_{\rm NL}$ and $n_{\rm NG}$ in our analysis, $V_{\rm surv}$ is the surveyed volume, and $f_{\rm sky}$ is the fraction of the sky observed. The integral over the momenta runs from $k_{\min}=2\pi/(V_{\rm surv}f_{\rm sky})^{1/3}$ to $k_{\max}=0.03h\rm Mpc^{-1}$, above which the non-Gaussian bias becomes negligible. Including shot-noise, the galaxy power spectrum $P_g(k)$ can be
\be
P_g(k, z)=b_{{\rm eff}}^{2}(k, z) P_{m}(k,z)+\frac{1}{\overline{n}}
\ee
where $\overline{n}$ is the mean number density of the survey. {In our analysis, {we assume the fiducial values $f_{\rm NL}=3$ and $n_{\rm NG}=0$} ruled by Planck \citep{akrami2018planck}}. We also use the minimal halo mass $M_{\min} = 10^{13}M_{\odot}$ and the pivot scale $k_p=0.016\rm Mpc^{-1}$. We have adopted the specification from Euclid \citep{laureijs2011euclid}, $z_{\rm median}=0.9$, $f_{\rm sky}=0.48$, $V_{\rm surv}=190 h^{-3}\rm{Gpc^3}$, together with the number density of $30$ galaxies per square arcminute. Finally, we obtain the standard deviations of the non-Gaussianity and its {non-Gaussianity's spectral index} {{are $\Delta n_{\rm NG}=1.74$ and $\Delta f_{\rm NL}=4.2$}. In the future, the LSS clustering data can also provide the complementarity constraining power on the {non-Gaussianity's spectral index}.

\section{Conclusions}
In this letter, {we analyze} the effects of the running of ``local'' type non-Gaussianity, originated from the general two scalar fields model, on mass function, large scale halo bias and correlation angular power spectrum comprehensively. The {non-Gaussianity's spectral index} $n_{\rm NG}$ {has effect} on the LSS clustering data at large scale. Therefore, we use the current LSS data: NVSS and SDSS DR6 QSOs as high redshit probes, together with SDSS DR12 PZs as lower redshit probe to compute their ACPS and CCPS with CMB map. Combining all data together, {we obtain the first} tight constraint on the {non-Gaussianity's spectral index} from the current LSS data: {$n_{\rm NG}=0.2 ^{+0.7}_{-1.0}$} ($1\,\sigma$ C.L.) at the pivot scale $k_p=0.016\rm Mpc^{-1}$. We also perform a forecast for $n_{\rm NG}$ using the future Euclid survey, {which shows} that the LSS clustering data can also useful on the estimation of the non-Gaussianity spectral's index of ``local'' type non-Gaussianity.

\section*{Acknowledgements}
J.-Q. Xia is supported by the National Science Foundation of China
under grants No. U1931202, 11633001, and 11690023; the National Key R\&D Program of
China No. 2017YFA0402600; the National Youth Thousand
Talents Program and the Fundamental Research Funds for the
Central Universities, grant No. 2017EYT01.





\bsp	
\label{lastpage}
\end{document}